\documentclass{article}

\usepackage[final]{nips_2016} 

\usepackage{graphicx}	
\usepackage{algorithmic}
\usepackage{amsmath}
\usepackage{amssymb}
\usepackage[ruled,vlined]{algorithm2e}
\usepackage{float}
\usepackage{geometry}
\usepackage[utf8]{inputenc} 
\usepackage[T1]{fontenc}    
\usepackage{url}            
\usepackage{booktabs}       
\usepackage{amsfonts}       
\usepackage{nicefrac}       
\usepackage{microtype}      
\usepackage[export]{adjustbox}
\usepackage[bottom]{footmisc}
\usepackage[utf8]{inputenc}
\usepackage[english]{babel}

\title{Finding Representative Points in Multivariate Data Using PCA}

\author{
Ashwinkumar Ganesan\\
      University Of Maryland Baltimore County (UMBC)\\
      Baltimore, MD, USA\\
      \texttt{mandarh1@umbc.edu}
\And
Tim Oates\\
      University Of Maryland Baltimore County (UMBC)\\
      Baltimore, MD, USA\\
      \texttt{gashwin1@umbc.edu}
\And
Matt Schmill\\
      University Of Maryland Baltimore County (UMBC)\\
      Baltimore, MD, USA\\
      \texttt{oates@cs.umbc.edu}
}

\begin{document}

\maketitle

\begin{abstract}
The idea of \textit{representation} has been used in various fields of study from data analysis to political science. In this paper, we define \textit{representativeness} and describe a method to isolate data points that can represent the entire data set. Also, we show how the minimum set of \textit{representative} data points can be generated. We use data from \textit{GLOBE} (a project to study the effects on Land Change based on a set of parameters that include temperature, forest cover, human population, atmospheric parameters and many other variables) to test \& validate the algorithm. \textit{Principal Component Analysis} (PCA) is used to reduce the dimensions of the multivariate data set, so that the \textit{representative} points can be generated efficiently and its \textit{Representativeness} has been compared against \textit{Random Sampling} of points from the data set.
\end{abstract}

\section{INTRODUCTION}

The study of environmental sciences and especially \emph{climate change} has grown in importance over the last two decades. Some reasons to study these areas are to understand their effects on human populations and on natural resources available on the planet. They encompass a host of issues that are related to diverse areas, including ecology, government policies, public health and economic policies \cite{GlobeProposal}. Global climate change study attempts to understand (and predict) how the climate evolves based on current and historical evidence. It explores the change in climate occurring due to natural phenomenon and anthropogenic reasons. Our effort is to find a way to globally utilize these studies by measuring similarity.  

\subsection{Land Change Science \& Globe}
Land Change Science (LCS) is a part of global climate change research and studies the effect of human activity on land and on the climate. As part of LCS, studies of specific geographic locations are conducted across the entire world. The locations where these studies are conducted are based on a number of factors that are decided depending upon the goals of the study that could be social or ecological. Goals that are social could include understanding the nature of human interaction with the environment and the impact of the change on human lives. Ecological studies try to measure parameters such as deforestation and its effect on the environment as a whole. The size or scale of a study could be based on the amount of funds available or to perform a study for measurement of global climate change (like an IPCC study) that can be used for making policy \cite{IPCC}. These studies are conducted by scientists on-site or by remotely monitoring these locations through stations that are constructed on the site.

\emph{Globe} is a global correlation engine to study land change science \cite{GlobeProposal}. It tries to fulfil the basic requirement of providing the ability to generalize LCS studies. It has multiple objectives which include constructing of an online social media system to facilitate collaborative work between land change scientists, providing a set of tools to analyse and generalize local observations to larger regions or the entire globe, and constructing scientific workflows \cite{GlobeProposal}.

\subsection{Motivation}
Land change studies are expensive. Hence the main motivation is to provide an understanding of how the conclusions of these studies can be generalized to other parts of the world. The notion of how generalized the studies are can help land change scientists to:
\begin{enumerate}
\item Reduce the number and cost of studies conducted in the future.
\item Select new locations in the world where studies have not been conducted.
\item Analyse existing studies in a different manner by showing a global pattern of the distribution of a set of selected parameters.
\end{enumerate}

\subsection{Representativeness}
There are a set of parameters or variables whose information is available for all parts of the world. They are the set of \emph{global} variables. These variables include potential vegetation, land utilized for agriculture, temperature, and many others. Scientists studying land change select a case study or a set of case studies which have been conducted. They select the variables which represent the environment in which the studies were conducted or the variables for which they want to analyze the case study. To generalize the results of these studies, they use these parameters to find other parts of the world that are \emph{similar}. The \emph{representativeness} of a given set of case studies is defined as the extent to which they \emph{cover} the rest of the world. For example, consider a scientist analyzing the Van Vliet Study \cite{vanVliet2012418} on trends in swidden cultivation. She would perform the following steps
\begin{enumerate}
\item Create a collection of all the locations that are part of study.
\item Select this \emph{collection} from \emph{Globe's} User Interface.
\item Select parameters such as \%crops, \%tree cover, temperature, Market Access Index and Potential Vegetation as the meta-study uses these variables.
\item Compute representativeness for the collection. Display it on a world map to show which other regions of the world are similar or to what extent the rest of regions are dissimilar to the ones in the study.
\end{enumerate} 

\subsection{Challenges \& Contributions}
The number of distinct geographical regions across the world into which the world map can be divided is very large. Each region is has its own global variable information. There are a large number of dimensions which, coupled with the number of sites, makes the representativeness computation expensive. The users (scientists in this case) select a subset of variables in realtime. The results for this calculation are to be provided in realtime. There are essentially two challenges here:
\begin{enumerate}
\item To find a method to measure representativeness.
\item To provide an algorithm, to find a minimum set of new locations where studies can be conducted so as to maximize representativeness for the selected set of variables.
\end{enumerate}

Our contribution in this thesis is to provide a mathematical formulation for representativeness, reduction of computation time and increase in efficiency by using a dimension reduction technique and providing a method to validate the results of our selected locations. Dimension reduction techniques are required since the number of dimensions is large and methods that use the original set of dimensions, such as clustering, will not be able to perform computation in realtime. The dimension reduction method we apply is Principal Component Analysis (PCA). PCA is useful because it prioritizes dimensions with higher variance. We show in this thesis how the algorithm can be used to select new locations and the correlation between the geographic points in the original space and single dimension PCA space. This correlation helps us measure the effectiveness of PCA and selected locations. Representativeness is displayed in the form of a \emph{Heat Map} using Google Maps. The map has markers that show the location where each case study was conducted.

\section{REPRESENTATIVENESS}
\subsection{Definition}
Representativeness describes how a data point or a set of data points can be used to generalize to the rest of the data set. In case of the Globe project, a data point refers to a specific region in the world where the study is conducted. Consider a distribution of data points where
\begin{itemize}
\renewcommand{\labelitemi}{$\bullet$}
\item \emph{D} is the given dataset of points
\item \emph{S} is a sample set of points such that \emph{S} $\subseteq$ \emph{D}
\item \emph{H} is a histogram based on \emph{D}
\item \emph{Bin(H,s)} is the bin where the data value \emph{s} falls in \emph{H}
\item \emph{P(H,i)} is the height / probability of bin \emph{i} in histogram \emph{H}
\item All unique bins are defined in a set
\emph{B = \{b$\vert$$\forall$$_{s \in S}$b = bin(H,s)\}}
\end{itemize}
We define \emph{representativeness R} of a sample set \emph{S} for a given global dataset \emph{D} as
\begin{equation} \label{eq:repequation}
\emph{R(S$\vert$D) = $\sum\limits_{b \in B} p(H,b)$}
\end{equation}
where 0 $\leq$ \emph{R(S $\vert$ D)} $\leq$ 1. When a sample set has higher representativeness, then \emph{R} reaches 1.

The definition of representativeness \emph{R} is theorized for a dataset \emph{D} which has a single variable or attribute for each data point. The histogram for the dataset \emph{D} gives us the frequency of data points in each bin that is defined. Once we know which data points fall in which bin and where points in sample set \emph{S} lie in the histogram, we know which data points are represented by \emph{S}. These data points are in set \emph{B} that is the set of bins \emph{b} where sample points in \emph{S} lie (i.e., $b \in B$). The data point \emph {x $\in$ D} is said to be represented by a sample point \emph{s $\in$ S}, when a certain pre-defined criteria is fulfilled.

Thus representativeness can be explained as a fraction of the total number of data points that fall within a predefined threshold criteria for atleast one of the points in the sample set. All data points within the threshold are completely represented by one of the sample points in \emph{S}. If a data point falls within the threshold criteria of multiple sample set points, then it is represented by the sample set point where criteria is optimal. In Globe, representativeness shows the fraction of the total land surface on earth that are similar to the locations that are part of a case study and have been studied based on a specific set of parameters.

If we use a multivariate dataset with \emph{m} dimensions, then the criteria used are modified to consider \emph{m} dimensions. For example, if the criteria is based on Euclidean Distance, then distance in a single dimension would be
\begin{equation} \label{eq:distancemetric}
\emph{d = $\vert$(s - x)$\vert$}
\end{equation}

where
\begin{itemize}
\renewcommand{\labelitemi}{$\bullet$}
\item \emph{d} is the distance
\item \emph{s} is a sample point such that \emph{s $\in$ S}
\item \emph{x} is a data point such that {x $\in$ D}
\end{itemize}

The distance formula for \emph{m}-dimensional data points would be
\begin{equation} \label{eq:ndimdistance}
\emph{d = $\sqrt{\sum\limits_{i \in m} (s_i - x_i)^2}$}
\end{equation}

The distance between the sample point \emph{$s \in S$} and \emph{x} shows how close the data point is to the sample point. As $d\rightarrow0$, the data point is considered to be closer. Representation of a data point by a sample point is inversely proportional to the distance. Hence, representation \emph{r} is defined as 
\begin{equation} \label{eq:representationdef}
\emph{$r(x \vert s) = \vert1 - d_x\vert$}
\end{equation}
\emph{r} is thus a value between 0 and 1 (and maybe greater than 1 in some outlier cases). A scale is created from 0 to 1 and the data points are assigned to each section of the scale (forming histogram \emph{H}). Representativeness \emph{R} is taken as the proportion of the total number of data points that are there in the first scale between 0 and 1. This is because the representativeness provided in the definition is for a binary scale where as represetativeness can be in degrees (like in the case of a heat map explained in the next chapter).

The histogram \emph{H} can be of 2 types: equal probability and equal area as shown in the diagram below where \emph{x} is a single variable and \emph{p(x)} is the probability distribution function (pdf). 

\begin{figure}[H]
\begin{center}
\includegraphics[scale=0.5]{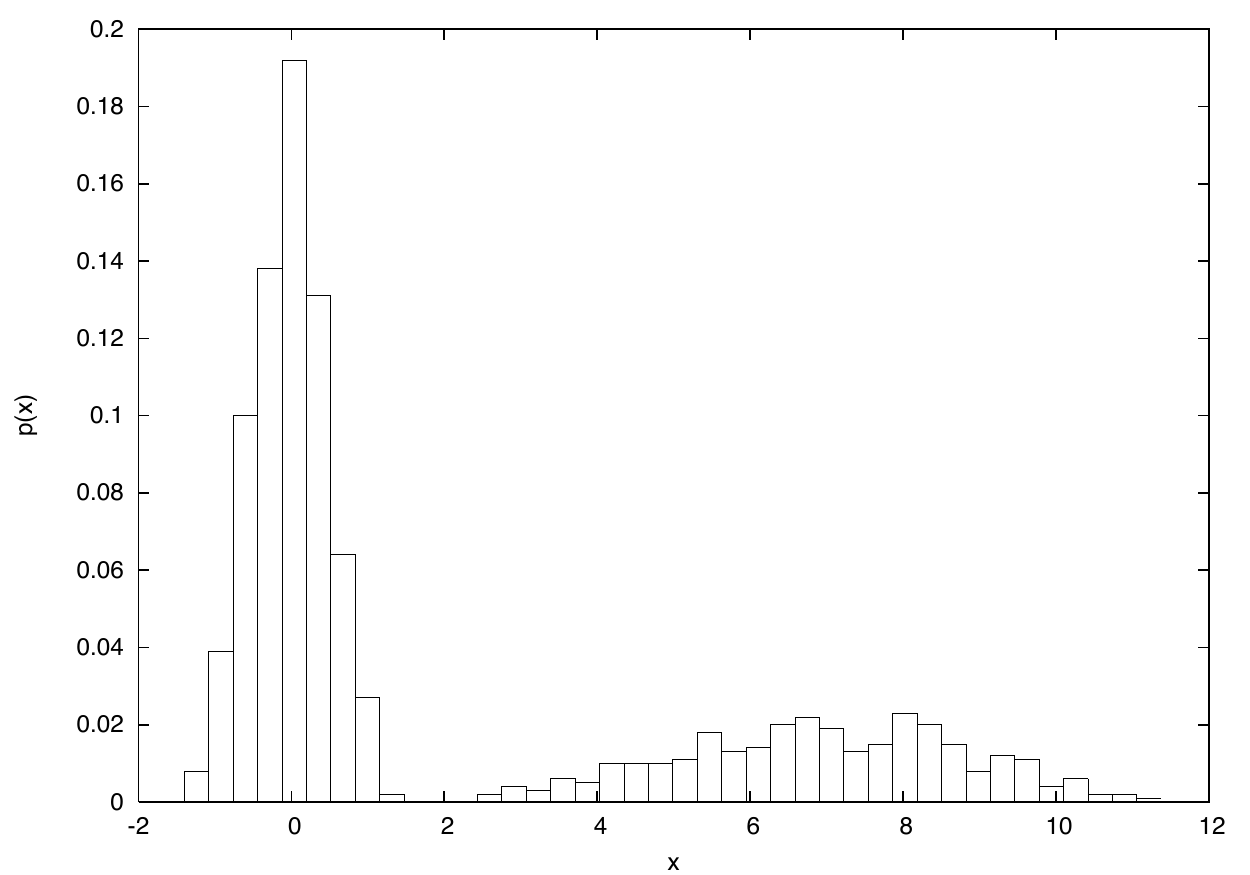}
\end{center}
\caption[\emph{Equal Area}]{\emph{A histogram of a sample of data from a distribution in which bins have equal area.}}
\label{fig:equiarea}
\end{figure}

\begin{figure}[H]
\begin{center}
\includegraphics[scale=0.5]{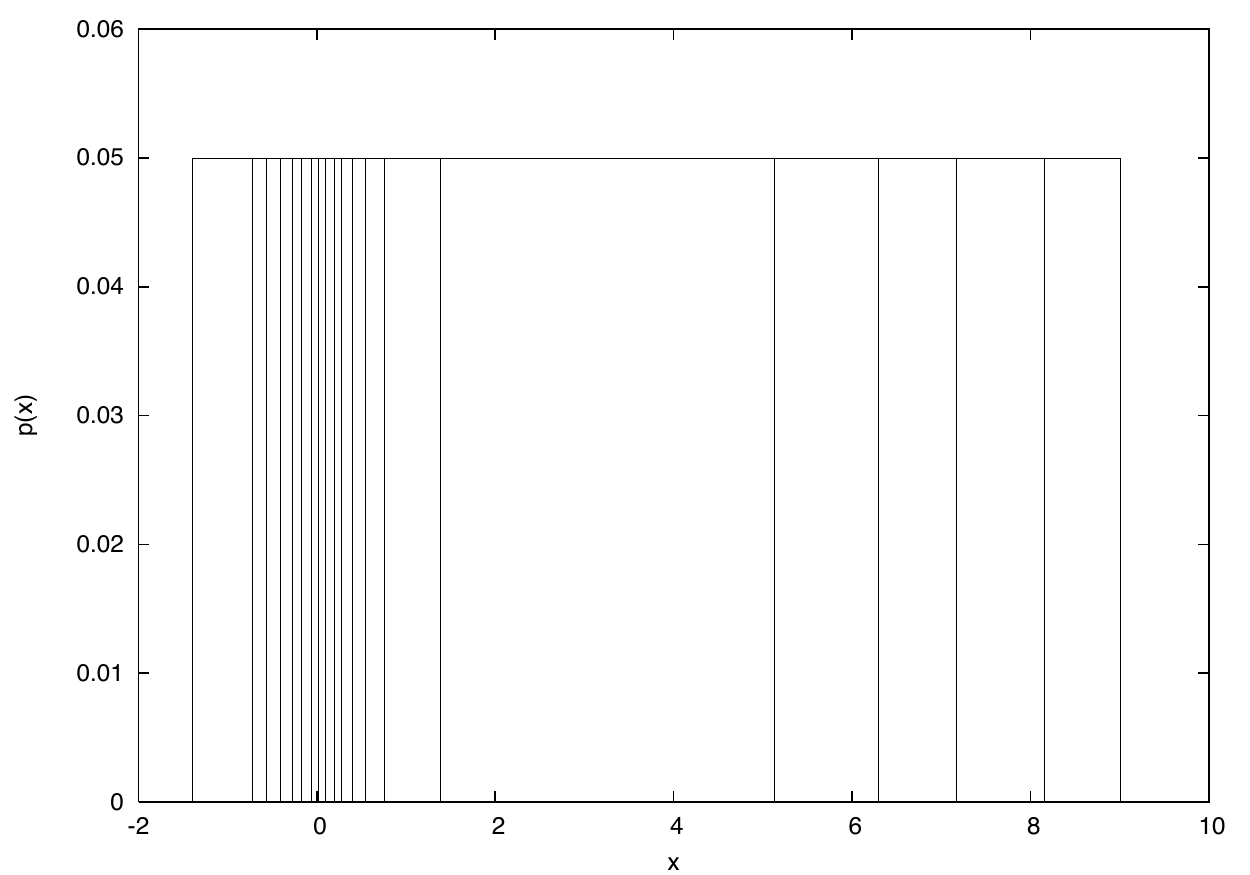}
\end{center}
\caption[\emph{Equal Probability}]{\emph{A histogram of a sample of data from a distribution in which bins have equal empirical probability.}}
\label{fig:equiprob}
\end{figure}

An equal area histogram is one where the data dimension is divided into bins of equal size (Figure \ref{fig:equiarea}). Thus each bin contains the points that fall within a certain range in the given dimension. Representativeness \emph{R} in such a histogram can be maximized by choosing mode points. An equal probability histogram is where the number of points in each bin is equal. Hence the size or width of each bin changes according to the density of the points (Figure \ref{fig:equiprob}). Since the bins are equiprobable, points can be selected from any random bin to represent the entire points in that bin.

\subsection{Kernels}
A kernel function is a function that maps a point onto a scale and is denoted by \emph{K(s - d)}. In the case of a histogram, the kernel function implemented is a step function. It is a set of bins whose points match the threshold criteria (like maximum distance) such that any point in any of the bins represents the all the other points. When the histogram is equiprobable, the step function is used to maximize the number of bins that are part of within the function's limits. The step function is shown in the diagram below.
\begin{figure}[H]
\begin{center}
\includegraphics[scale=0.5]{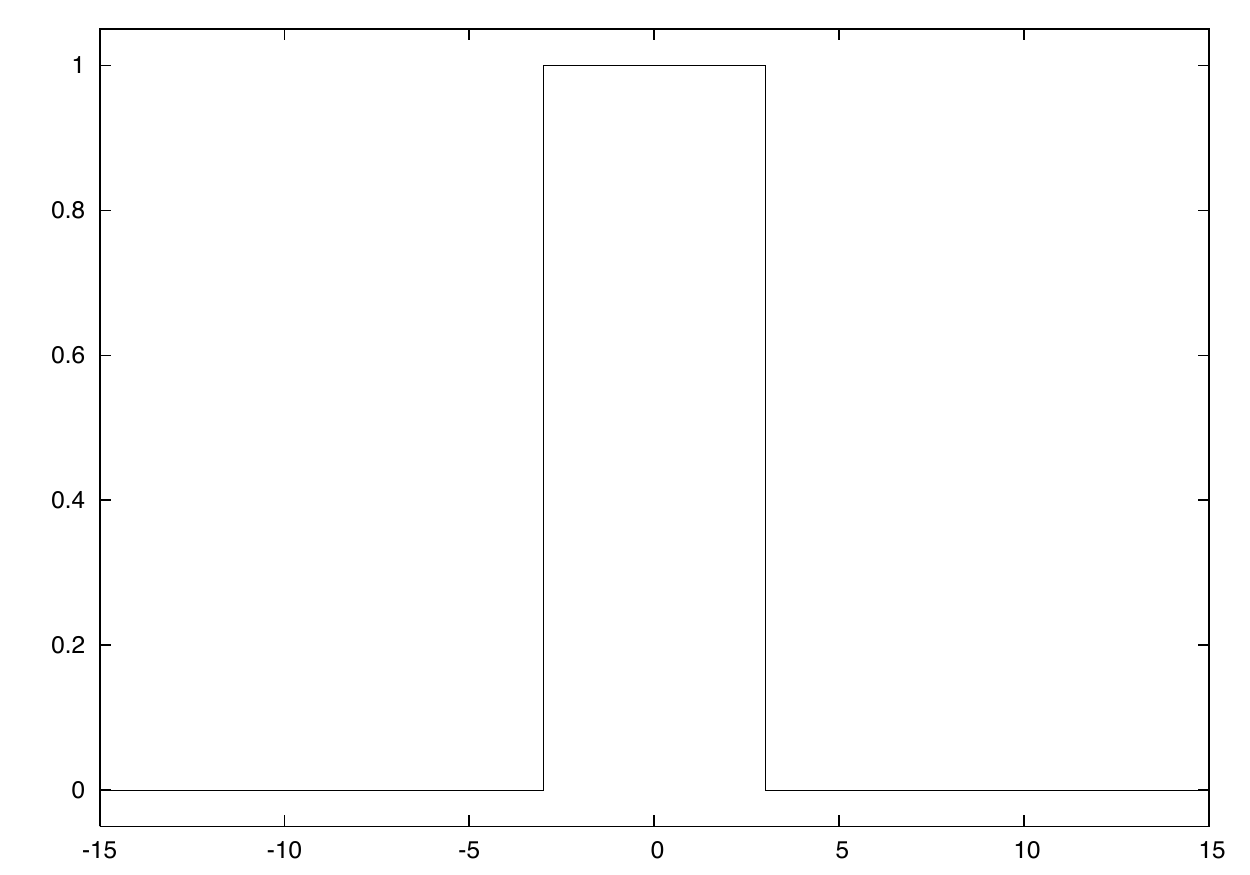}
\end{center}
\caption[\emph{Step Kernel}]{\emph{Sample Step kernel function.}}
\label{fig:stepkernel}
\end{figure}

\section{RELATED WORK}
In this section, we discuss various methods that can be used to maximize represetativeness. The main aim of the methods described below is to find an optimal set of sites or points such that representativeness can be maximized.

\subsection{Clustering}
Clustering techniques are a set of methods to group data points that are similar together \cite{Jain:1999:DCR:331499.331504}. These characteristics of the groups are defined by a pattern of values in their variables. Clustering is an unsupervised learning method. It does not require a training data set to create a model. The groups in which the data points are to be classified need not be known at the start. Hence, clustering can be used for exploratory data analysis to identify patterns in the data. Clustering is a three stage process
\begin{enumerate}
\item Extract features from the given set of points.
\item Perform similarity measurement between data points.
\item Create groups based on the similarity measurement.
\end{enumerate} 

Clustering techniques are of different types and mainly divided into 2 categories:-\\
\emph{Hierarchial Clustering Techniques} - These techniques create groups of points which are similar to each other. Once a group is formed, it creates the next level by combining groups that are similar. In this way, a hierarchy of groups is created with all groups merged at the top most level of the hierarchy. The structure is called a dendogram \cite{Jain:1999:DCR:331499.331504}.\\
\emph{Partitional Clustering Techniques} - These techniques try to create a single partition in the dataset as compared to a dendogram which may have a high computation time. The problem occurs when the size of the dataset is large. Partitional techniques try to optimize a certain function based on which the partition is made. Calculating the optimal set of values for the function could again be computationally expensive. Hence an approximation is calculated by executing the algorithm multiple times on the same dataset until the function reaches a state that is \emph{close} to optimal. For example, using squared error a as function to create partitions \cite{Jain:1999:DCR:331499.331504}. The algorithm is executed until the squared error is reduced to a value that is below a certain pre-determined threshold.

\subsection[K-Means Clustering]{K-Means Clustering}

The k-means clustering algorithm is a widely used algorithm \cite{Vattani:2009:KRE:1542362.1542419}. This is a centroid or partition based clustering technique. The algorithm clusters all the data points into \emph{k} clusters. The algorithm starts by selecting an arbitrary set of centroids $c_1, c_2 ... c_k$. It then assigns each point to the closest centroid $c_i$. Once the points are clustered, it calculates the center of mass for each cluster to get a new set of centroids. The previous steps are then repeated for the new centroids. After each iteration the set of centroids moves closer to the final set such that the next iteration does not change the set of centroids chosen. This means the center of mass for the \emph{k} clusters calculated remains constant. The algorithm stops computing after this point. The worst case time complexity is $O(n^{\emph{kd}})$ \cite{Vattani:2009:KRE:1542362.1542419} where \emph{n} is the number of data points, \emph{k} is the number of clusters and the points are in a \emph{d}-dimensional space. 

\subsection[Nearest Neighbor Clustering]{Nearest Neighbor Clustering}
This is a hierarchial clustering technique. In this clustering method, the nearest neighbor to each data point is found and the point is assigned to that cluster. A Voronoi decomposition of the data points is performed \cite{Bubeck_nearestneighbor}. There is a threshold or quality function $Q_n$ to put a threshold on the distance that is considered between the point and the cluster. Thus all the points are put into \emph{k} clusters where \emph{k} is user-defined. The clustering is implemented using a graph based structure. Whenever a point closest to the current point is found, an edge is created between them thus linking them in the same cluster \cite{Jain:1999:DCR:331499.331504}. It is also called agglomerative single-link clustering technique and has a time complexity of $O(n^2)$ \cite{Manning:2008:IIR:1394399}.

K-means clustering and nearest neighbor clustering can be used to find a set of \emph{k} centroids that maximize representativeness. K-means clustering generates \emph{k} clusters with unique centroids that are the representative points. For nearest neighbor clustering, we can select any point randomly from each of the \emph{k} clusters generated (as all clusters adhere to the quality function $Q_n$), as they represent the other points within the cluster.

\subsection{Dimension Reduction Techniques}
Consider a data set where each point has a large number of variables. These variables may have different scales of values, and different densities and variances. There are a number of possible problems with high dimensional data \cite{DRReview}:
\begin{enumerate}
\item Processing high dimensional data (especially when the number of data points is large) is expensive.
\item Even though the number of dimensions is high, the data could be classified or clustered using a smaller subset of variables.
\item As the number of dimensions increases, the values for some variables may become sparse. This is known as the \emph{empty space problem} \cite{DRReview}.
\item The \emph{Curse of Dimensionality} states that the number of sample points required to approximate a function increases exponentially as the number of variables / dimensions increases. 
\end{enumerate}

A dimension reduction technique is a transformation which reduces number of dimensions required to represent a sample. The reduced set of dimensions may be a subset of the original set of dimensions (for example, using information gain) or could be a completely new set of  dimensions. Some of standard dimension reduction techniques that can be used to transform a high dimensional data set are Principal Component Analysis (PCA), and Self Organizing Maps (SOM)\cite{DRReview}. Neural Networks with GIS have also been used for constructing a \emph{Land Transformation Model} which tries to forecast how usage changes \cite{Pijanowski2002553}. Self Organizing Maps have been used to perform environmental assessment of regions, grouping based on environmental conditions, and finding out which areas might deteriorate in the future \cite{springerlink:10.1007/s00267-003-2917-6}. Once a SOM is trained, the \emph{k} nodes from the weight vector can be used as centroids representing their respective clusters. As the nodes may not be actual data points, the point closest to each node will be used as a representative point. Training a SOM may require updating the weight vector over several iterations of the data set. The time complexity of a SOM is $O(\vert E \vert \bullet (\vert E \vert + \vert V \vert))$\cite{4436179}.

\subsection{Clustering Using a Combination Of Methods}
Hoffman et. al.\cite{MSTCRep} use \emph{Multivariate Spatio-Temporal Clustering (MSTC)} to calculate representativeness of sampling networks. MSTC can be performed using a combination of PCA and K-Means clustering\cite{MSTCRep}. The data set considered is high dimensional and is assumed to contain a lot of redundant information. Hence the method involves reducing the number of dimensions using PCA at the beginning and then performing standard k-means clustering. Hoffman et. al. also provide a set of improvements for performing PCA and k-means clustering. The time required to perform k-means clustering is reduced by decreasing the number of distance computations between the centroid and the other points, based on cluster created and new distances computed. The time complexity of PCA computation is reduced by parallelizing it. The summation of all euclidean distances from points to their nearest sample locations or centroids, is used to measure representativeness of the sample set. Higher the sum, lower is the representativeness of the sample set.

\section{CALCULATING REPRESENTATIVENESS}
\subsection{Principal Component Analysis (PCA)}
Principal Component Analysis (PCA) is a technique used to analyze multivariate datasets and to find patterns in the data. It is used as a dimension reduction technique where higher dimension data points can be projected onto a lower dimension space \cite{WICS:WICS101}. It takes a data set of \emph{n} dimensions and projects it onto a set of new dimensions (which again can be a maximum of \emph{n}) that are orthogonal to each other (although all \emph{n} need not be used). The new dimensions are not correlated with each other. They are called \emph{Principal Components}. The principal components are calculated by performing an eigen decomposition of the covariance matrix. The covariance between dimensions in the original dataset is used because those dimensions which have a higher variance, provide more information regarding the nature of the dataset \cite{reDRtech}. Another reason to use PCA is because it is easy to understand and efficient algorithms exist to compute each step \cite{reDRtech}. It is also a widely used method because it is able to reduce noise and the dimension reduction performs data clustering \cite{Ding:2004:KMC:1015330.1015408}. As we shall see, the cost of computing distance is also reduced as the final distance is in a single dimension.

\subsection{Measuring Representativeness Using PCA \& Histograms}
The following section provides the details of the algorithm to calculate Representativeness of a given sample set of regions and how to select a new set of sites to improve representativeness across the entire world for the selected set of attributes or variables. Land Change scientists know many regions across the world are not represented by sample sites in their case studies by design. In \emph{Globe}, they are provided the option to filter out these regions, so that they are not considered as part of the analysis. Hence we apply our method to the list of unfiltered regions.

The following diagram shows the algorithm's workflow:
\begin{figure}[H]
\centering
\includegraphics[width=75mm]{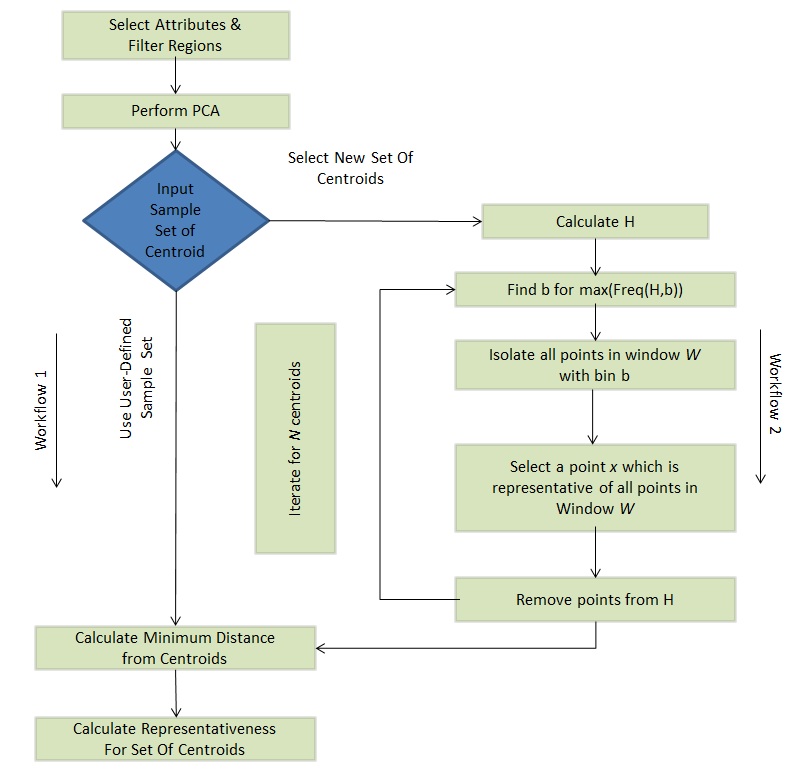}
\caption{\emph{Workflow to Calculate Representativeness}}
\end{figure}

As shown in the diagram above, the algorithm has two distinct workflows to solve two objectives:
\begin{enumerate}
\item The first workflow calculates the representativeness for the user-defined Sample Set of regions where case studies were conducted.
\item The second workflow provides a method to select \emph{N} regions to maximize representativeness for the user-defined set of attributes. These are the places where the system can recommend that a study could be conducted. Once the regions are selected the representativeness of these sites is calculated.
\end{enumerate}

Both workflows use a distance metric to calculate how close a location is to another location where a study was conducted.
\begin{equation} \label{eq:distance}
\emph{FD(x) = min($D_p(x) - D_p(s)_{s \in S}$)}
\end{equation}
where \emph{x} is a specific location across the world and \emph{s} is a place in the world where a study was conducted. \emph{s} is part of a larger Sample Set \emph{S}. \emph{$D_p$} is the projection of a location (or data point) onto the first principal component. The final distance \emph{FD(x)} is the minimum distance between \emph{x} and \emph{S}.

Representativeness of the Sample Set, as given in equation \ref{eq:repequation} can now be transformed to the following equation:
\begin{equation} \label{eq:rep}
\emph{R(S $\vert$ D) = $\frac{\sum\limits_{x \in C} \sum\limits_{b \in W} Freq(H,b)}{D}$}
\end{equation}
where \emph{R(S $\vert$ D)} is the representativeness of Sample Set \emph{S} and \emph{D} is the complete dataset. \emph{H} is the histogram of \emph{$D_p$}. \emph{Freq(H, b)} is the number of points in (or the frequency of) any bucket \emph{b} in histogram \emph{H}. The window size \emph{W} that is a set of buckets \emph{b}. A selected region represents all regions in \emph{W}. \emph{W} can be user-defined. \emph{C} is the set of regions or centroids selected.

\subsubsection{Preconditions}
Before any of the following workflows are executed, the user selects a subset of the global variables (such as temperature, \%tree cover) for which representativeness is measured. Also, the user is allowed to filter out a set of regions from the dataset. For example, the user may choose to limit the dataset only to regions which are in the tropics (i.e., having a temperature between $15^{o}C$ and $31^{o}C$). Once the data is filtered, we get a final list of regions and matrix containing the value of each user selected variable for each of these regions.

\subsubsection{Workflow 1 - Calculate representativeness of given sample set}
The steps to calculate representativeness of a sample set of sites (equation \ref{eq:rep}) that are provided by the user are:\\

\textbf{Step 1}

Select the a set of variables based on which representativeness is measured. Calculate the eigenvalues and eigenvectors for all the unfiltered regions across the world. If the unfiltered regions do not contain the sites that are part of the Sample Set, then find the first principal component and project all the regions (including the sites) onto this dimension.\\

\textbf{Step 2}

Calculate the distance between each region and a region in the Sample Set. The Final Distance is the minimum distance found. It shows which location in the Sample Set is closest to the current location being considered and how close it is.\\

\textbf{Step 3}

\begin{figure}[H] \label{fig:colorscheme}
\centering
\includegraphics[width=75mm]{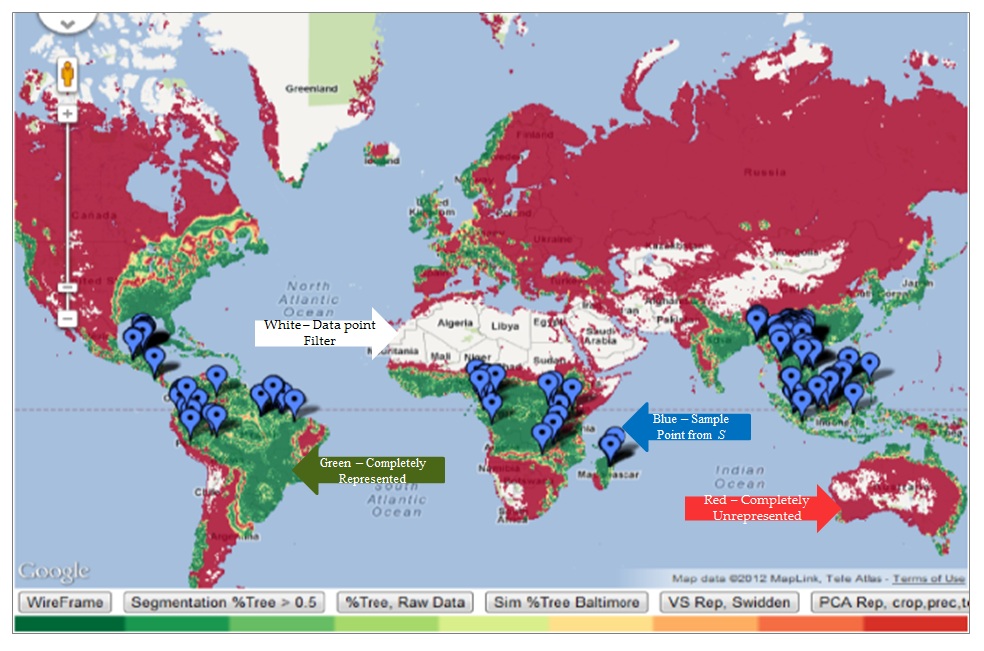}
\caption{\emph{Representativeness Coloring Scheme}}
\end{figure}

Display all the locations on Google Maps with  a color chosen according to the final distance calculated. All the distance are a value between 0 \& 1. A color pallete consists of a set of shades (e.g. from Green to Red where Green is considered as a place completely represented by one of the sample sites while Red is completely unrepresented). The scale between 0 \& 1 is divided based on the number of shades which are in the color pallete. The color associated with the bucket in which the final distance of a place falls, is applied on Google Maps, on top of the location of that place. The following diagram (figure \ref{fig:colorscheme}) is an example of how the color scheme looks.

\subsubsection{Workflow 2 - Creating a set of \emph{Ideal} sites}
An \emph{ideal} set of sites is a set of sites that maximizes representativeness but at the same time has a minimum number of sites required to reach that measure of representativeness. The steps to find a set of \emph{ideal} set of sites that are representative of all other regions in the world are:\\

\textbf{Step 1}

As seen in the previous workflow, calculate the eigenvalues and eigenvectors for all the unfiltered regions across the world.\\

\textbf{Step 2}

Once the regions are projected onto the first principal component, create the histogram of that dimension. The histogram requires a number of buckets into which the dimension is divided. The number of buckets is defined after testing. Once the histogram is constructed, a window of size \emph{W} is defined such that 
\begin{center}
\emph{1 $\leq$ \emph{W} $\leq$ number of buckets} 
\end{center}

The window is a set of buckets (containing regions) such that any region in the bucket represents all the regions in the buckets which are present in \emph{W}. \emph{W} is either user-defined or an arbitrarily fixed size. The algorithm performs the following steps \emph{N} times where \emph{N} is the number representative sites to be found.

\begin{algorithm}[!ht]
\DontPrintSemicolon
\KwIn{A finite set $FB_P=\{fb_1, fb_2, \ldots, fb_m\}$ of Frequency of buckets in Histogram $H$, window size $W$}
\KwOut{A finite set $C$ containing a set of $N$ representative Points}
\emph{rand()} is picks a random point from a bucket $b_k$\\
\emph{WUsed} is bit array of size $\lfloor\frac{B_P}{W}\rfloor$\\
$B_P$ contains points each bucket. $B_P=\{b_1, b_2, \ldots, b_m\}$\\
$MW$ is the window with Maximum Frequency.\\
\For{$i \gets 1$ \textbf{to} $\lfloor\frac{B_P}{W}\rfloor$}{
\emph{WUsed}$_i \gets false$
}
\For{$i \gets 1$ \textbf{to} $N$}{
 $max \gets -1$\;
 $j \gets 1$\;
 $MW \gets -1$\;
 $maxBin \gets -1$\;
 \While{$k \leq n$} {
   \If{$(fb_k \geq max)$ \&\& $($\emph{WUsed}$_j = false)$} {
      $max \gets fb_k$\;
      $maxBin \gets b_k$\;
      $MW \gets j$\;
     }
     \If{$mod(k,W) = 0$ \&\& $(k/W) \geq 0$ }{
       $j \gets j+1$\;
     }
     $k \gets k+1$\;
   }
   \emph{WUsed}$_{MW} \gets true$\;
   $P \gets rand(b_{maxBin})$\;
   $C_i \gets P$\;
}
\Return{$C$}\;
\caption{Finds $M$ points to maximize Representativeness of Points}
\label{algo:selectp}
\end{algorithm}

Algorithm \ref{algo:selectp} shows how an \emph{ideal} set of sites is selected. Consider a histogram \emph{H}, with window size \emph{W}. Let $B_P$ contains points each bucket. $MW$ is the window with the maximum frequency. $\lfloor\frac{B_P}{W}\rfloor$ is the total number of windows in histogram \emph{H}. We maximize representativeness by selecting a single point from a window. Once a window is utilized, it not used again. This is because a point from a given window of buckets represents all the points in the window completely. \emph{WUsed} is an array of bits that shows which of the windows have used.  \emph{WUsed} is initialized to $false$. Then, the algorithm iterates through each bucket in the histogram and finds the bucket which has the maximum frequency (mode) and that is part of a window that has not been used before (i.e., where  \emph{WUsed}$_i$ is $false$) . A bucket $b_k$ is part of a window $\lfloor(k/W)\rfloor$. If bucket $b_k$ is the bucket with the maximum freqeuncy, then make \emph{WUsed}$_{\lfloor(k/W)\rfloor}$ as $true$.  Then, select any point from $b_k$ and add it to the final list of centroids \emph{C}. The above set of steps performed to find a centroid, is repeated \emph{N} times to get a set \emph{C} containing \emph{N} centroids.

The algorithm has to take care of a specific condition i.e. the case when the number of centroids required is higher as compared to the number of bins that are greater than 0. In such a case, the number of centroids returned is limited to number of non-zero bins that can be found.\\

\textbf{Step 3}\\
This step is the same as the previous workflow, where the regions selected are displayed on a World Map with the representativeness of the entire set.

\section{EXPERIMENTAL RESULTS \& ANALYSIS}
Land Change scientists also compare their sample sites against random sampling, as a standard practice. Each experiment conducts 3 sets of tests. The first test is to calculate the representativeness of the sample sites given by the user. The second test uses the \emph{histogram} method discussed in the previous chapter to generate a new set of sites (the number of sites is equal to that defined by the user) where a study can be conducted. It also calculates the representativeness for the same. The third test generates representativeness for a random set of sites for the same number of sites as in the previous two cases. Random sampling is performed 1000 times to eliminate any bias created from a limited set of random sampling tests. We define the ideal sample set as the set of samples generated by the histogram method \ref{algo:selectp}. 

\subsection{Measuring Representativeness}
The Van Vliet study has been used for the purpose of conducting experiments \cite{vanVliet2012418}. This meta-study conducts a global assessment of swidden cultivation i.e. slash and burn. It is an agricultural technique where forested areas are burned to create fields for agriculture. There are a total of 157 sites that are part of the study. These sites are used as centroids in our experiment to measure its representativeness. The window size for all experiments below is 1 and number of bins in the histogram is 157.

\begin{figure}[H]
\begin{center}
\includegraphics[scale=0.4]{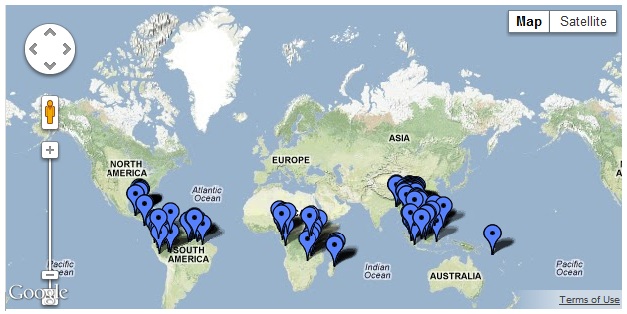}
\end{center}
\caption[\emph{Sites in Van Vliet Study}]{\emph{Sites in Van Vliet Study}}
\label{fig:vvsites}
\end{figure}

Figure \ref{fig:vvsites} shows the location  of sites that are part of the Van Vliet Study. 

\subsubsection{Measuring with a Filter}
The following parameters are applied in the experiment:\\
\textit{Filter} is applied so that regions of the world the author does not claim to represent are not considered in the analysis. In the example, the parameter \emph{potential vegetation} is used to filter the data set. Potential vegetation has a range of values form 0 to 12 \cite{Ramankutty}. The values considered in the experiment are from 1 to 2. These values are used filter out all regions except the tropical regions (and some forested areas) across the world.\\\\
A total of 3 attributes or variables are used in the experiment viz. \emph{potential vegetation}, \emph{market access} and \emph{temperature}. The selection of the variables is based on the study.\\\\
\textit{Zoom Level} decides the total number of points which are considered in the dataset. For example, at zoom level 4, the total number of points or regions is 160,000. The experiments are conducted at zoom level 6. The total number of points at this level is about 1.4 million. When the filter is applied, the total number of \emph{Global land units GLU's} is reduced to about 250,000.\\\\
\textit{Color Scale} applied has a total of 10 colors from red to green. It depicts the different levels of representation in the \emph{heat map}.\\

\begin{figure}[H]
\begin{center}
\includegraphics[scale=0.35]{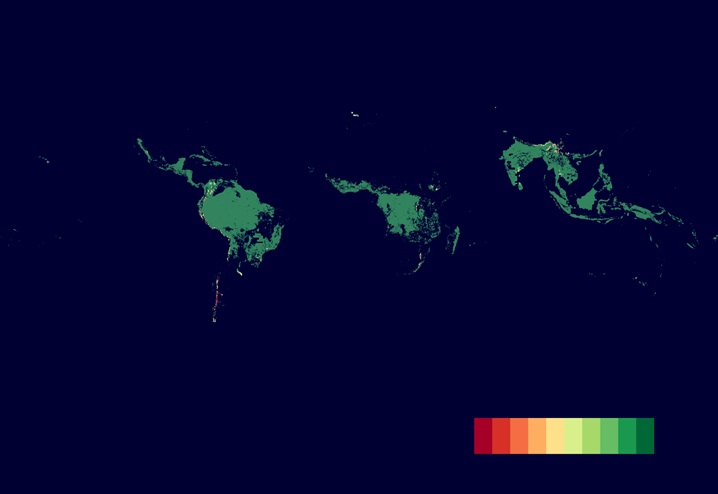}
\end{center}
\caption[\emph{Representativeness of Filtered World Regions for Given Samples}]{\emph{Representativeness of Filtered World Regions Using Given Samples}}
\label{fig:repwrgs}
\end{figure}

\begin{figure}[H]
\begin{center}
\includegraphics[scale=0.35]{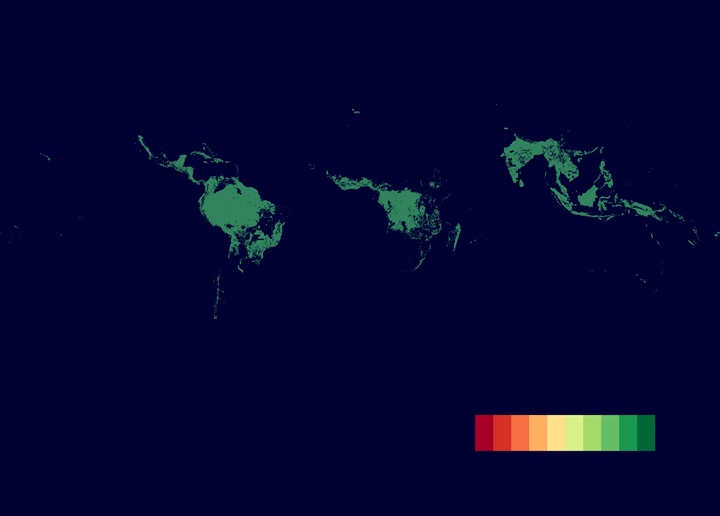}
\end{center}
\caption[\emph{Representativeness of Filtered World Regions for Ideal Samples}]{\emph{Representativeness of Filtered World Regions Using Ideal Samples}}
\label{fig:repwris}
\end{figure}

\begin{figure}[H]
\begin{center}
\includegraphics[scale=0.35]{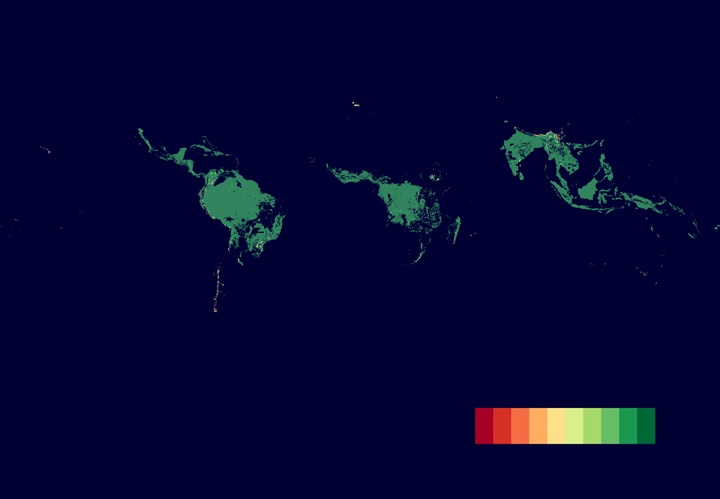}
\end{center}
\caption[\emph{Representativeness of Filtered World Regions for Random Samples}]{\emph{Representativeness of Filtered World Regions Using Random Samples}}
\label{fig:repwrrs}
\end{figure}

The figures \ref{fig:repwrgs}, \ref{fig:repwris}, and \ref{fig:repwrrs} depict the \emph{heat map} generated for the Van Vliet study. The color scale is provided at the bottom of each image. The scale goes from red to green, where green depicts ``complete represention" and red depicts ``complete non-representation". The parts of the world that are filtered out are shown in dark blue. This contains all the water bodies and the other regions of the world that do not fit the filter criteria.

\begin{figure}[H]
\begin{center}
\includegraphics[scale=0.7]{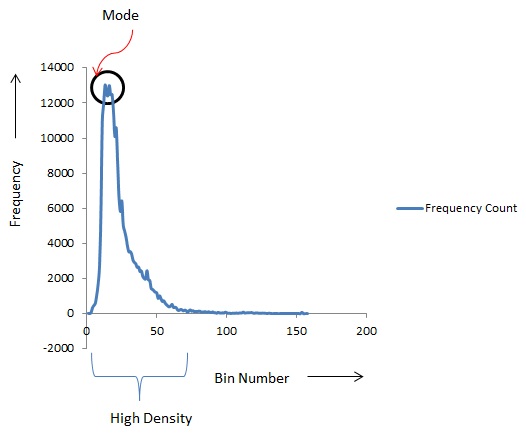}
\end{center}
\caption[\emph{Histogram Of First Principal Component Values (Filtered)}]{\emph{Histogram Of First Principal Component Values (For Filtered Data) of 1.4 Million Regions}}
\label{fig:filterpcahistogram}
\end{figure}

Figure \ref{fig:repwrgs} shows some regions that are not represented by the locations in the study. The maps generated using given, random and ideal samples are almost the same. This is because of the definition of representativeness. To maximize representativeness, the sampled regions have to ``cover" as much of the filtered regions as possible. To represent a set of regions that have similar conditions, only a single point or location is required. Based on the filter used, the potential vegetation, temperature and market access parameters are of similar areas as can be seen on the map. This makes regions cluster with high frequency in specific regions in the PCA space, thus making the number of locations required to represent them lower than 157. This is seen in the histogram in figure \ref{fig:filterpcahistogram} where the high frequency bins require only a single location in the bin to represent the entire bin. Thus random sampling and the histogram method to calculate new locations have representativeness values that are close to \emph{1.0} or $100\%$. 

\begin{figure}[H]
\begin{center}
\includegraphics[scale=0.35]{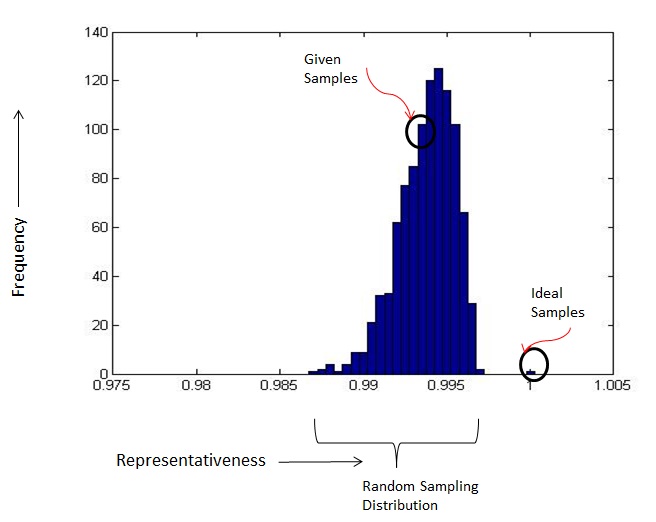}
\end{center}
\caption[\emph{Representativeness Distribution}]{\emph{Histogram Of Representativeness for Random Sampling and Where Other Methods Lie}}
\label{fig:histrepfilter}
\end{figure}

Figure \ref{fig:histrepfilter} shows the comparison of representativeness of the given sample and the ideal samples generated against random sampling conducted. Representativeness for the given sample set at $48^{th}$ percentile while the ideal sample set is at 100 percentile. It means that the given sample set is better than random sampling only 48\% of the time as compared to ideal sample set which is better 100\% of the time. This makes the ideal sample set a better selection as compared to random or the given sample set. Table \ref{tab:frep} shows that representativeness of the given sample set is better the representativeness of the average random sample set.

\begin{figure}[H]
\begin{center}
\includegraphics[scale=0.5]{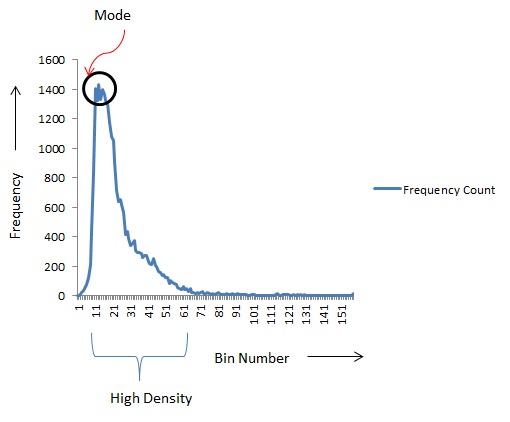}
\end{center}
\caption[\emph{Histogram Of First Principal Component Values (Filtered)}]{\emph{Histogram Of First Principal Component Values (For Filtered Data) of 160000 Regions}}
\label{fig:filterpcahistogramlower}
\end{figure}

\begin{table}[H]
\centering
\begin{tabular}{| c | c|}
  \hline
  \bf{Method} & \bf{Representativeness} \\
  \cline{1-2}
  Given Sample & 0.994 \\
  \hline
   Ideal Sample & 1.0 \\
  \hline
  Avg. Random Sampling & 0.9937 \\
  \cline{1-2}
\end{tabular}
\caption{\emph{Representativeness Of Samples for Filtered Data}}
\label{tab:frep}
\end{table}

Figure \ref{fig:filterpcahistogramlower} shows that the same histogram trend is maintained when we perform PCA on a downsized dataset. Hence the calculations made on a ISea3H level 10 are applied to the ISea3H level 12 hexagons. Thus representativeness, based on a  downsized number of regions, does not change as nature of distribution remains the same. The total number of regions for the downsized set after filtering is 27883.

\subsubsection{Measuring without a Filter}

This section describes an experiment to compare the various samples without a filter being applied. The data set is the complete 1.3 million regions across the globe. The world maps generated show a clearer differentiation in the given sample representativeness against that of an ideal and a random sample. 

\begin{figure}[H]
\begin{center}
\includegraphics[scale=0.35]{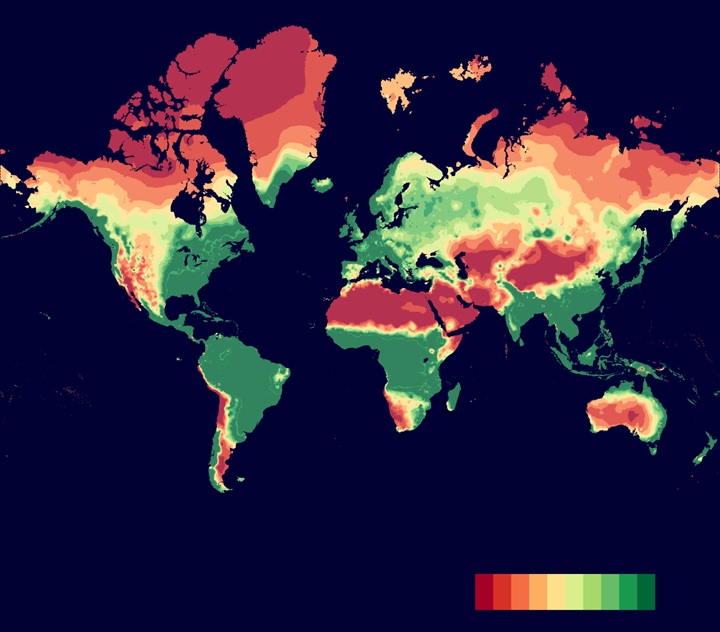}
\end{center}
\caption[\emph{Representativeness of Unfiltered World Regions for Given Samples}]{\emph{Representativeness of Unfiltered World Regions Using Given Samples}}
\label{fig:repwnrgs}
\end{figure}

\begin{figure}[H]
\begin{center}
\includegraphics[scale=0.35]{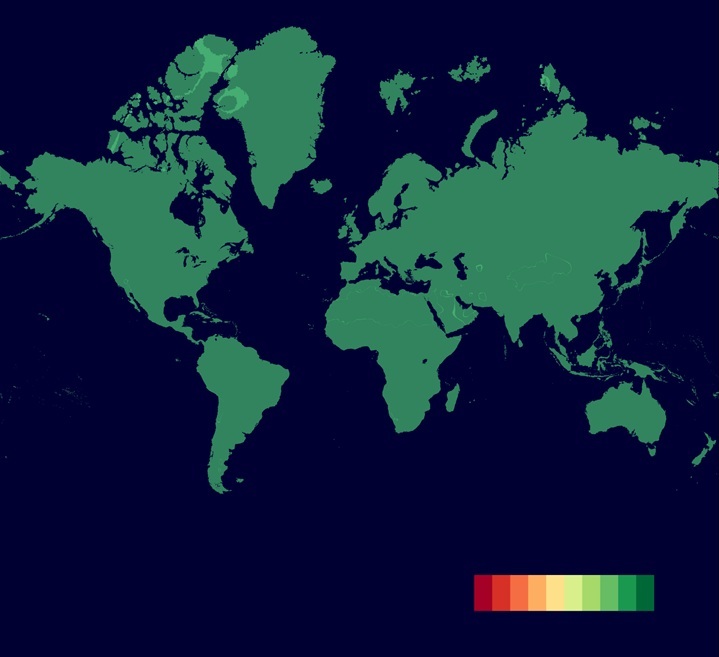}
\end{center}
\caption[\emph{Representativeness of Unfiltered World Regions for Ideal Samples}]{\emph{Representativeness of Unfiltered World Regions Using Ideal Samples}}
\label{fig:repwnris}
\end{figure}

\begin{figure}[H]
\begin{center}
\includegraphics[scale=0.35]{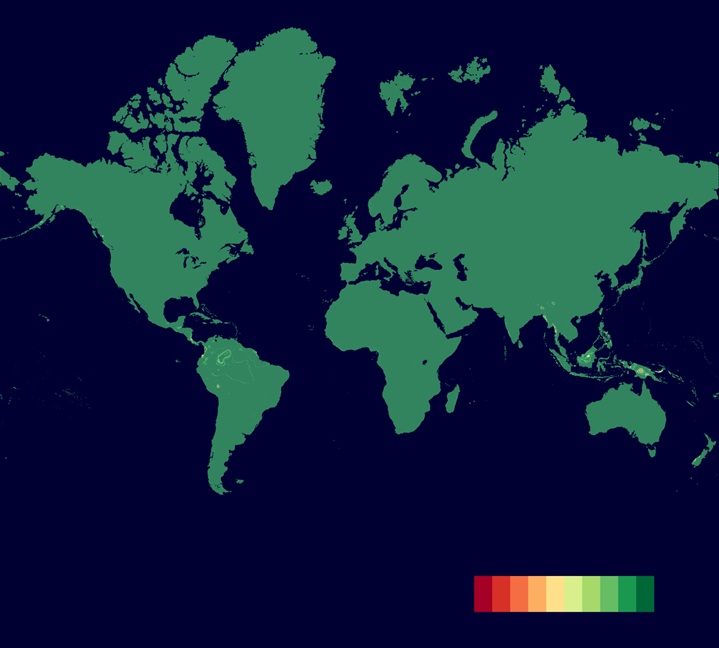}
\end{center}
\caption[\emph{Representativeness of Unfiltered World Regions for Randoml Samples}]{\emph{Representativeness of Unfiltered World Regions Using Random Samples}}
\label{fig:repwnrrs}
\end{figure}

\begin{figure}[H]
\begin{center}
\includegraphics[scale=0.30]{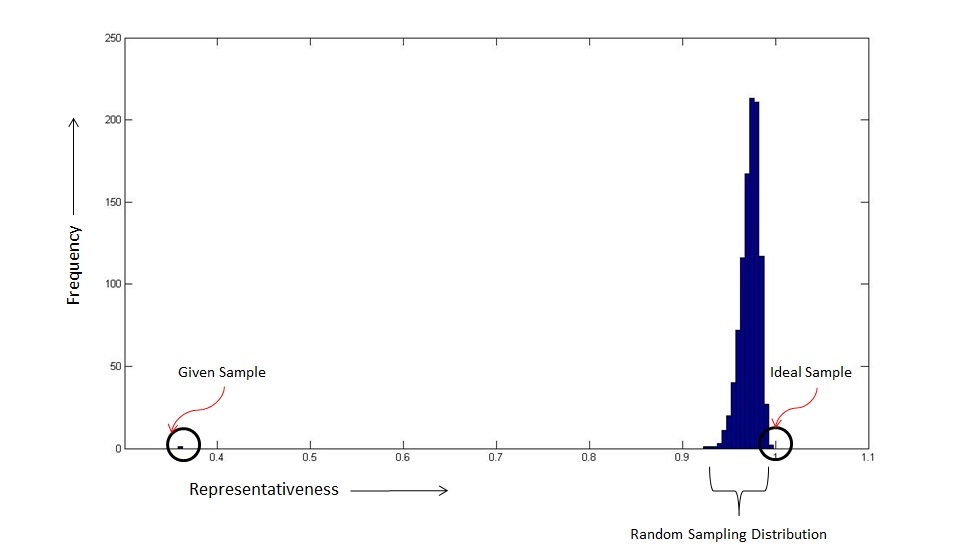}
\end{center}
\caption[\emph{Representativeness Distribution}]{\emph{Histogram Of Representativeness for Random Sampling and Where Other Methods Lie}}
\label{fig:histrepnf}
\end{figure}

Figure \ref{fig:histrepnf} gives an idea of the distribution of representativeness values for random sampling and shows where the given sample and ideal samples are placed in the distribution. Table \ref{tab:nfrep} shows that ideal sampling has the highest representativeness.

\begin{table}[H]
\centering
\begin{tabular}{| c | c|}
  \hline
\bf{Method} & \bf{Representativeness} \\
  \cline{1-2}
  Given Sample & 0.362 \\
  \hline
   Ideal Sample & 0.995 \\
  \hline
  Avg. Random Sampling & 0.972 \\
  \cline{1-2}
\end{tabular}
\caption{\emph{Representativeness Of Samples for Unfiltered Data}}
\label{tab:nfrep}
\end{table}

The variables selected affect the nature of the distribution. A uniform distribution of points across variables will reduce the amount variation covered by a single principal component. For example, if the variables distributed the data points in the form of a circle (in a 2 dimensional graph), then there would not be a one principal component that would cover a large portion of the variation limiting the use of PCA. The variables we have used in our experiments have a large variation (e.g. temperature) where as potential vegetation is a categorical value limiting variation.

\subsection{Ideal vs Random Sampling}
In the previous section, the results show that ideal sampling has a representativeness close to 1.0, the same as random sampling. This result, as described previously, is a function of the number of sites (or centroids) that need to be generated. Since the number of sites (157) is large, the representativeness is close to \emph{1.0}. We can analyze the effectiveness of ideal sampling by measuring the representativeness of ideal samples against random sampling for a reduced of sample size. We consider the unfiltered data points as in previous example. The following graph shows the trend for representativeness for the two methods, starting from the selection of 1 sample site to a total of 157.

\begin{figure}[H]
\begin{center}
\includegraphics[scale=0.4]{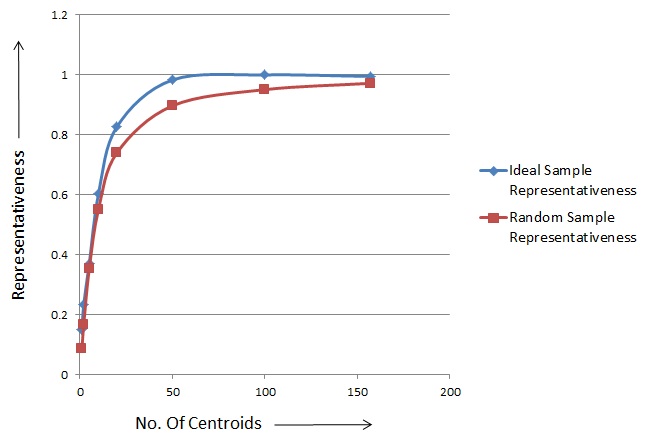}
\end{center}
\caption[\emph{Representativeness Trend - Increasing Number Of Centroids}]{\emph{Representativeness Trend - Increasing Number Of Centroids}}
\label{fig:reptrendnf}
\end{figure}

The trend shows that ideal sample representativeness tends to 1 to with fewer centroids as compared to random sampling, making it better to select sites as compared to random sampling. Table \ref{tab:nfcrep} shows the number of centroids required by each method to reach a value for representativeness close to 1.

\begin{table}[H]
\centering
\begin{tabular}{| c | c | c|}
  \hline
\bf{Method} & \bf{Centroids} & \bf{R} \\
  \hline
  Ideal & 60 & 0.99 \\
  \hline
  Avg. Of Random & 100 - 130 & 0.95 - 0.967 \\
  \hline
\end{tabular}
\caption{\emph{Number Of Centroids \& Representativeness}}
\label{tab:nfcrep}
\end{table}

\subsection{Measuring the Effect of Histogram Size}
\begin{figure}[H]
\begin{center}
\includegraphics[scale=0.35]{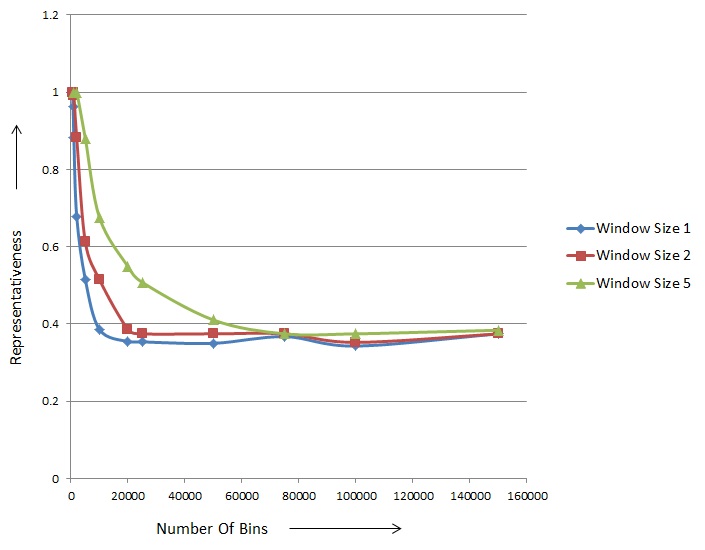}
\end{center}
\caption[\emph{Representativeness Trend - Increasing Number Of Bins}]{\emph{Representativeness Trend - Increasing Number Of Bins}}
\label{fig:nfbinvsrep}
\end{figure}

The ideal samples are generated by constructing a histogram of the first principal component. The algorithm \ref{algo:selectp} is affected by 2 parameters, i.e., the number of bins created in the histogram and the window size. The window size is the number bins in which any data point can be considered to represent the rest of the data points with the bins in the window completely. Currently, the window size and number of bins are chosen after testing. Figure \ref{fig:nfbinvsrep} shows the change in representativeness for various window sizes and varying number of bins.

The test has been conducted for window sizes 1, 2 \& 5. The minimum number of bins taken is equal to the number of centroids required which is maintained at 157 throughout all the tests. As the window size is 1, there are a total of 157 bins from which a data point can be selected (refer to the algorithm \ref{algo:selectp}). As the number of bins is increased, the representativeness decreases, until it reaches a set of bins after which the representativeness becomes stable. The second observation is that the rate of decrease in representativeness is lower when the window size is increased. Thus a higher window size can reduce rate of decrease in representativeness. We explain the reasons for the decrease in the next section.

\subsubsection{Effect of the Number of Bins}
\begin{figure}[H]
\begin{center}
\includegraphics[scale=0.5]{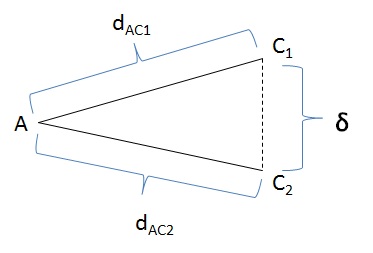}
\end{center}
\caption[\emph{Example Of Bin Distance}]{\emph{Example for Bin Distance}}
\label{fig:examplebin}
\end{figure}

Consider figure \ref{fig:examplebin}. There are 3 points $A, C_1,$ \emph{and} $C_2$,  where $C_1  \&  C_2$ are 2 centroids. $d_{AC1}  \&  d_{AC2}$ are the distances between the points $A  \&  C_1$ and $A  \&  C_2$ in the first principal component respectively. $\delta$ is the distance between $C_1  \&  C_2$. From the diagram, we see if $\delta \rightarrow 0$, then $(d_{AC1} - d_{AC2}) \rightarrow 0$. 

\begin{equation} \label{eq:interval}
\emph{IntervalSize = $\frac{p_{max} - p_{min}}{NoOfBins}$}
\end{equation}
where $p_{max}$  and  $p_{min}$ are the maxmimum and minimum values in the first principal component respectively. 

Thus, when the number of bins increases, IntervalSize decreases. Consider the case when window size is 1. Consider figure \ref{fig:filterpcahistogram} where there are certain parts of the histogram that have high density. As the IntervalSize decreases, centroids are chosen from bins adjacent to the mode where $\delta \rightarrow 0$. The final representativeness is based on the color scale that is used. All the data points that fall in the first bin of the scale are considered completely represented by the sample set. As more centroids are selected, the difference in the distance either remains the same or get smaller. The redundant centroids generated are unable to \emph{cover} the rest of the points in the distribution to minimize the distance and maximize coverage. Thus,

\begin{equation} \label{eq:proportion}
\emph{NoOfBins $\alpha$  Colorscale}
\end{equation}

Representativeness stabilizes after a certain number of bins because the histogram is divided into small parts such that coverage is only for the high density part of the histogram from where all the centroids are selected. In such a case, increasing the number of bins no longer affects the representativeness.

\subsubsection{Effect of the Window Size}
The window size creates a minimum distance between any 2 centroids that are selected as only a single point can be selected from within a bin in a certain window at a time. Hence figure \ref{fig:nfbinvsrep} shows a lower rate of decrease in representativeness as the number of bins increase. For a optimal representativeness, the ideal sample set selected should be function of the number of bins and size of the window used. Hence equation \ref{eq:proportion} can be modified to,

\begin{equation} \label{eq:newproportion}
\emph{$\frac{NoOfBins}{WindowSize}$  $\alpha$  $Colorscale$}
\end{equation}

\section{CONCLUSION \& FUTURE WORK}
We have provided a definition and an algorithm for calculating the representativeness of a set of sample sites. When the number of dimensions increases, clustering methods become computationnally inefficient. This is specifically when the representativeness needs to be calculated in near realtime. Hence dimension reduction techniques are used. We have used principal component analysis to perform n-dimension reduction into a single dimension based on the variance of attributes. This helps us project points onto the first principal component with maximum difference or spacing between points. To nullify the effect of the magnitude of the values in each attribute, normalization of each column is performed. This makes all the values fall in the scale from 0 to 1. The distances from the given sample set to other points are calculated using their first principal component projected values. We see that the results of the heat map generated to show representativeness across the globe is as expected. To maximize representativeness, we have provided a method that is based on the creating a histogram of the PCA values and selecting modes. The method is able to maximize representativeness as seen in the experiments conducted. The samples drawn from land change literature and the ideal samples are compared against random sampling. We show that if a filter is applied, the given sample set has the lowest representativeness as compared to the other two methods (figure \ref{fig:histrepfilter} and figure \ref{fig:histrepnf}), even though representativeness is greater than 0.9. It also shows that ideal sampling is better than random sampling. We are also able to see, as shown in figure \ref{fig:reptrendnf}, that ideal sampling reaches the same measure of representativeness as compared to random sampling with fewer samples. Hence it is better than random sampling at performing site selection. We also analyze the properties of the ideal sampling method for each of the parameters affecting the method, mainly the number of bins in the histogram and the window size. We show (figure \ref{fig:nfbinvsrep}) that when the number of bins is increased the representativeness decreases until it reaches a stable level. 

In the future, a set of improvements can be performed. These are:
\begin{enumerate}
\item Creating a function correlation between the number of bins in the histogram to the scale applied for representativeness. The current method applies an arbitrary number of bins in the histogram. This helps to maximize representativeness under all conditions.
\item The scale applied is a linear scale from 0 to 1 that is divided equally. There is no correlation between the scale and the actual distance calculated between the centroids and the other points. If the initial projected values are very small, then the distances calculated are also small. Thus a region that is not related to any of the centroids can be shown as being represented. One solution to this problem is to normalize the projected values again, so that in case the values are very small they are scaled up accordingly to a value between 0 and 1. The points that have a value greater than 1 can be considered as outliers. 
\item The problem of the distance (between a sample and another region in PCA space) being very small may still exist. The representativeness scale can be changed to account for regions in the principal component where the density of points is concentrated. Thus, density estimation and PCA outlier detection can be performed to create a tighter lower and upper bound of projected data points. 
\item The visualization of the map can be improved by implementing isolines or contour lines. Contour lines or isolines are lines across which the function has the same output value. Thus isolines can be implemented for all the areas that have the same projected value on the first principal component.
\end{enumerate}

\bibliography{main}
\end{document}